# Model-based Decentralized Bayesian Algorithm for Distributed Compressed Sensing

Razieh Torkamani[a,*], Hadi Zayyani[b], Ramazan Ali Sadeghzadeh[a]

[a] Faculty of Electrical Engineering, K.N. Toosi University of Technology, Tehran, Iran, 16317-14191

[b] Faculty of Electrical and Computer Engineering, Qom University of Technology (QUT), Qom, Iran, 37195-1519

*Corresponding author.

E-mail addresses: rtorkamani@mail.kntu.ac.ir (R. Torkamani), zayyani@qut.ac.ir (H. Zayyani), sadeghz@eetd.kntu.ac.ir (R. A. Sadeghzadeh).

**Abstract:**

In this paper, a novel model-based distributed compressive sensing (DCS) algorithm is proposed. DCS exploits the inter-signal correlations and has the capability to jointly recover multiple sparse signals. Proposed approach is a Bayesian decentralized algorithm which uses the type 1 joint sparsity model (JSM-1) and exploits the intra-signal correlations, as well as the inter-signal correlations. Compared to the conventional DCS algorithm, which only exploit the joint sparsity of the signals, the proposed approach takes the intra- and inter-scale dependencies among the wavelet coefficients into account to enable the utilization of the individual signal structure. Furthermore, the Bessel K-form (BKF) is used as the prior distribution which has a sharper peak at zero and heavier tails than the Gaussian distribution. The variational Bayesian (VB) inference is employed to perform the posterior distributions and acquire a closed-form solution for model parameters. Simulation results demonstrate that the proposed algorithm have good recovery performance in comparison with state-of the-art techniques.

## 1. Introduction

Compressive sensing (CS) has attracted much attention in recent years [1-6]. The CS theory develops a novel approach for simultaneous sampling and compressing of sparse signals. In CS, linear combinations of the signal samples are measured and the number of required measurements is considerably smaller than the number of the samples acquired by Shannon/Nyquist theorem [1]. According to the CS, the measurements are converted to an underdetermined set as the following formula and the goal is finding the sparsest solution of this underdetermined equation set:

$$y = Ax + n \tag{1}$$

where $y \in \mathbb{R}^{M \times 1}$ is the vector of CS measurements, $A \in \mathbb{R}^{M \times N}$ denotes the measurement matrix, $x \in \mathbb{R}^{N \times 1}$ is the vector of sparse coefficients, and $n \in \mathbb{R}^{M \times 1}$ is the noise of measuring process.

Finding the sparsest solution of the above underdetermined equation set corresponds to solving the following optimization problem:

$$\min \|x\|_0 \quad s.t. \quad y = Ax \tag{2}$$

where $\|.\|_0$ is the zero norm and denotes the number of nonzero elements of the vector. Since $M < N$, (2) is an NP-hard problem [2]. One method for solving (1) is the sparsity promoting scheme, such as:

$$\tilde{x} = \arg\min_{x} \left\{ \frac{1}{2} \|y - Ax\|_2^2 + \lambda \|x\|_p \right\} \tag{3}$$

where $\|.\|_p$ is the $\ell_p$-norm with $p \in [0,1]$, and $\lambda > 0$ is a penalty parameter to balance the importance of the reconstruction error and the sparse prior. For $p = 1$, the expression in (3) corresponds to the least absolute shrinkage and selection operator (LASSO) or basis pursuit denoising (BPDN) algorithm [7, 8]; for $p = 0$, it becomes the iterative hard thresholding (IHT) algorithm [9]; for $p \in (0,1)$ it becomes the iterative reweighted algorithm [10].

Another approach for reconstructing sparse signal from CS measurements is the Bayesian CS (BCS) or sparse Bayesian learning (SBL) algorithm [11, 12], which generally provides better performance. SBL posits a prior probability for variables and utilizes Bayesian inference to estimate the unknowns.

However, CS only exploits intra-signal dependencies, i.e. the sparsity of the single signal, and ignores the multi-signal correlations. In order to exploit the intra- and inter-signal correlations simultaneously, distributed CS (DCS) is proposed in [13, 14], which is an extension of CS and is usable for jointly sparse signals. Two or more signals are said to be jointly sparse if they possess common nonzero support set, i.e. their nonzero coefficients belong to common index set. The main framework of DCS includes independently compression of each signal and then, jointly reconstruction of multiple sparse signals. DCS has been successfully employed in many applications such as wireless sensor networks (WSNs) [15, 16], video coding [17, 18], image fusion [19, 20], multiple-input multiple-output (MIMO) channel [21], multichannel electroencephalogram (EEG) [22], joint channel estimation [23], ground moving target indication (GMTI) [24], and multi-channel SAR system [25]. Three joint sparsity models (JSMs) are presented in [13]: JSM-1, JSM-2, and JSM-3 model. In this paper we apply JSM-1 model to the joint sparse signals, which assumes a common component and an innovation component for each individual signal.

Recovery algorithms for DCS are divided into two categories: centralized and decentralized approaches. In a centralized manner, all the CS measurements acquired independently at each node is gathered in a fusion center (FC) and then, the FC estimates multiple-sparse signals jointly. The estimated signals are sent back to their nodes. In contrast, in a decentralized approach, each node accomplishes its own sparse signal estimation by authorizing intercommunication between neighboring nodes, in the absence of a fusion center.

Since the theory of DCS is exposed to discussion, many algorithms have been proposed for solving joint recovery problem of the typical models (JSM-1, JSM-2, and JSM-3). In [13], Duarte et al. propose one-step greedy algorithm (OSGA) for joint reconstruction of the source signals. The OSGA considers JSM-2 model and estimates the multiple sparse signals in an iterative greedy manner. In [15], a decentralized algorithm based on sparse Bayesian learning method is proposed. In this work, the JSM-1 DCS model is used and variational Bayesian (VB) inference is exploited to solve the DCS recovery problem. In [26], a weighted $\ell_1$-norm minimization based on JSM-1 model is used and a single linear program is exploited for joint sparse signal reconstruction. In [27], a distributed parallel pursuit (DIPP) algorithm is developed, in which the sensors communicate information about the estimated support-sets. The exchange of information helps to improve estimation of the partial common support-set.

In all the distributed recovery algorithms, a dependence model is employed to signify the intra- and inter-signal correlations between multiple sparse signals. While this demonstrates significant progress from conventional CS, our argument in this work is that it is possible to achieve even better results by introducing a model-based DCS algorithm.

In this paper, we aim to go beyond simple joint sparsity by exploiting the coefficients interdependency structure among the elements of each individual signal. To achieve this end, every multiscale transform can be used as sparsifying basis, such as wavelet, curvelet, contourlet, shearlet transforms. In this work, the discrete wavelet transform (DWT) is applied as the multiscale sparse domain. Also, an accurate prior is used which exploits the inter- and intra-scale dependencies of the DWT coefficients. The inter-scale dependency among the DWT coefficients is implied by the hidden Markov tree model (HMT), and the intra-scale dependency is modeled by a second-order neighborhood system.

The most widely used statistical model in the CS literature, and specifically, the only model that is used in the DCS literature is the Gaussian pdf. In the work presented here a Bessel-k form (BKF) pdf is proposed as a suitable prior to impose the sparsity of the signals. The motivation for using the BKF pdf is its sharp peak at zero, which leads to its superiority in modeling the intra-scale correlations of the DWT coefficients over the Gaussian pdf [28, 29]. Also, it has a heavier tail than the Gaussian pdf.

To the best of our knowledge, a BKF pdf for modeling the sparse signal coefficients was used in [29]. Also the prior model employed in this paper for the support vector has already been used by [30]. However, the approaches in [29] and [30] are developed for single sparse signal and the jointly reconstruction of multiple-sparse signals have not been considered there. Moreover, Markov chain Monte Carlo (MCMC) inference was employed there to derive the posterior distributions, which is not computationally efficient and the convergence is difficult to assess. In the proposed algorithm, which we call the Wavelet-based Bayesian DCS algorithm based on BKF prior (WBDCS-BKF), we propose a modified model to that in [29] and [30], which uses VB inference procedure, and hence, the closed form solutions are completely different from those obtained by [29] and [30]. Also, a decentralized algorithm is proposed to jointly recover sparse signals in DCS, which follows the JSM-1 DCS setting. In recent years, a simple and powerful algorithm to solve the decentralized optimization problems, the alternating direction method of multipliers (ADMM), is developed [31, 32]. In this paper, the ADMM technique is applied to solve the decentralized VB algorithm.

Full names and abbreviations used in paper are presented in Table1.

The rest of this paper is organized as follows: Section 2 describes the wavelet-based BCS formulation using the BKF prior; In Section 3, we provide the proposed DCS algorithm; The VB inference procedure for the centralized and decentralized approaches are developed in Sections 4 and 5, respectively; Experimental results are presented in Section 6; and Section 7 concludes the paper.

**Table 1:** Full names and abbreviations used in paper.

| Full Name | Abbreviation |
| --- | --- |
| Alternating Direction Method of Multipliers | ADMM |
| Bayesian Compressive Sensing | BCS |
| Bessel K-form | BKF |
| Basis Pursuit Denoising | BPDN |
| compressive sensing | CS |
| distributed compressive sensing | DCS |
| Backtracking-Based Adaptive Orthogonal Matching Pursuit for Block DCS | DCSBBAOMP |
| Discrete Cosine Transform | DCT |
| Discrete Wavelet Transform | DWT |
| Electroencephalogram | EEG |
| Fusion Center | FC |
| Ground Moving Target Indication | GMTI |
| Hidden Markov Tree | HMT |
| Iterative Hard Thresholding | IHT |
| joint sparsity model | JSM |
| Least Absolute Shrinkage and Selection Operator | LASSO |
| Markov Chain Monte Carlo | MCMC |
| Multiple-Input Multiple-Output | MIMO |
| Normalized Mean Square Error | NMSE |
| Orthogonal Matching Pursuit | OMP |
| One-Step Greedy Algorithm | OSGA |
| Peak Signal to Noise Ratio | PSNR |
| Synthetic Apperture Radar | SAR |
| Sparse Bayesian Learning | SBL |
| Tree-Structured Wavelet-based CS using MCMC | TSWCS-MCMC |
| Variational Bayesian | VB |
| Wavelet-based Bayesian DCS algorithm based on BKF prior | WBDCS-BKF |
| Wireless Sensor Network | WSN |

## 2. Wavelet-based BCS using BKF prior

In this section, we explain the CS reconstruction via Bayesian framework and use BKF model as a suitable prior for sparse coefficients.

### 2.1 Bessel-K Form distribution

A popular prior for DWT coefficients is the Gaussian distribution [11, 13]. Nevertheless, it has been proved that the BKF distribution captures the heavy tail behavior of DWT coefficients better than the Gaussian model [29]. The BKF pdf is represented by

$$p_x(x|p,c) = \frac{1}{\sqrt{\pi}\Gamma(p)} \left(\frac{c}{2}\right)^{-\frac{p}{2}-\frac{1}{4}} \left|\frac{x}{2}\right|^{p-\frac{1}{2}} K_{p-\frac{1}{2}}\left(\sqrt{\frac{2}{c}}|x|\right) \qquad (4)$$

where $K_v(.)$ is the modified Bessel function of the second kind of order $v$, $p > 0$ is the scale parameter, $c > 0$ is the shape parameter, and $\Gamma(v) = \int_0^\infty z^{v-1} e^{-z} dz$ is the Gamma function.

The BKF pdf is unimodal and symmetric around the mode [33]. The peakedness of the pdf relies on the value of the parameter $p$, such that any increase in $p$ leads to an increase in the peakedness of the pdf. For $p = 1$, the BKF pdf simplifies to the double exponential PDF. If $p > 1$, it approaches to the Gaussian distribution, especially when $p \gg 1$. The range $p < 1$ is of special interest, because the tails of the BKF pdf become heavier and the peak becomes sharper, and thus, it can model the non-Gaussian statistics of the wavelet coefficients [28].

Proposition 1: Let $w$ be a random variable with zero-mean Gaussian distribution, i.e. $p(w|\tau) = \mathcal{N}(0,\tau)$, where $\tau$ is the variance of the pdf and has Gamma distribution: $p(\tau) = Gamma(a,b)$, where $a,b > 0$. In this case, the resulting unconditional distribution for $w$ is $p(w) = BKF(a, \frac{a}{b^2})$ [34].

This hierarchical prior model is referred to as a doubly stochastic process and characterize the correlations between adjacent wavelet coefficients. In this paper, we utilize this prior model in our proposed algorithm.

### 2.2 Wavelet-based BCS using BKF prior

Here, we develop the wavelet-based BCS recovery problem using BKF prior model. Suppose that $x \in \mathbb{R}^{N \times 1}$ is the original signal. The 2-D DWT of the signal $x$ can be expressed as [35]

$$x = \Psi\theta \qquad (5)$$

where $\Psi$ is an $N \times N$ matrix containing the DWT basis vectors as its columns, and $\theta \in \mathbb{R}^{N \times 1}$ is the vector of DWT coefficients. The DWT coefficients can be represented in a tree structure. According to the wavelet-tree structure, each coefficient at scale $s$ is a parent for four children at the next scale, and the statistical dependency of the parent-children relationship is such that the children of the small-valued parent are also small [35]. However, if the parent coefficient is large, then its children may be large or small. This inter-scale relationship between the wavelet coefficients can be demonstrated via the hidden Markov tree (HMT) model [36]. Moreover, the intra-scale dependency of the wavelet coefficients is specified by the value of the neighbors of each coefficient, such that if the number of negligible coefficients in the neighboring block is higher than a threshold, then the value of the current coefficient is likely to be also small. Fig.1 presents a three level DWT decomposition of a 2-D signal, in which the neighbors of two coefficients at scale $s = 3$ and two wavelet trees have been highlighted. As it is obvious, the neighboring coefficients should be specified among the elements within the same scale. In this paper, we consider second-order neighborhood system, i.e. $3 \times 3$ neighboring block for exploiting intra-scale correlation.

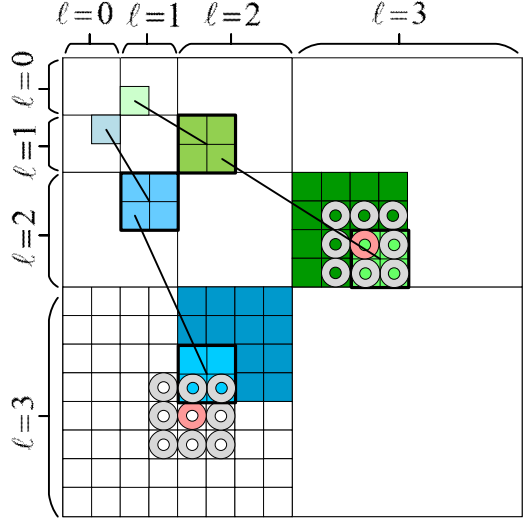

**Fig. 1:** Wavelet coefficients structure with tree structure depicted across scales and neighbors of two coefficients.

It has been proved that most of the signals and images have sparse representation in the wavelet domain [35], thus, enabling us to utilize the CS theory. In this paper, we develop the Bayesian formulation for CS inversion problem. Within the Bayesian framework, we estimate the sparse coefficients vector from measurements via imposing a suitable prior belief for hidden variables.

According to the "Central Limit Theorem", the measurement noise $n$ follows a zero-mean Gaussian pdf with variance $\lambda_n$ [11]. Thus, we have the following statistical model

$$p(y|\theta, z) = \mathcal{N}(Ax, \lambda_n I_N) = \mathcal{N}(D\theta, \lambda_n I_N) \qquad (6)$$

where $D = A\Psi$ is the $M \times N$ measurement matrix, and $I_N \in \mathbb{R}^{N \times N}$ is an identity matrix. In this paper, we assume a BKF prior distribution for the sparse wavelet coefficients

$$\theta = w \odot z \qquad (7)$$

$$p(w) = \prod_{i=1}^{N} p(w_i^s) \qquad p(w_i^s) = \text{BKF}(a, b) \qquad (8)$$

where $w \in \mathbb{R}^N$ is the signal coefficients, $z \in \mathbb{R}^N$ is the support vector of the sparse signal, $\odot$ denotes the Hadamard product, and $w_i^s$ is the $i$'th component of signal $w$, which is located at scale $s$.

## 3. Bayesian DCS Model

The main goal of this paper is implementing the HMT model for representing statistical dependencies between the locations of the individual sparse signal coefficients signal and, then, developing an algorithm for jointly reconstruction of multiple correlated signals. In this section, we develop the proposed WBDCS-BKF algorithm by extending the BKF- based BCS algorithm presented in the previous section and following JSM-1 DCS model for modeling joint sparse signals.

Let K be the number of network nodes. We can model the network by an undirected graph $\mathcal{G} = \{\mathcal{V}, \mathcal{E}\}$, where $\mathcal{V} = \{1, ..., K\}$ is the set of nodes and $\mathcal{E} \subset \mathcal{V} \times \mathcal{V}$ is the set of undirected edges and depicts the links between the nodes. In a specific graph, two nodes are said to be neighbor if there is a link between them to communicate information. Fig. 2 displays an example graph with 7 nodes.

As mentioned in the introduction, each node performs CS measurements independently, which is given by

$$y_k = D_k \theta_k + n_k \qquad k = 1, ..., K \tag{9}$$

where $y_k \in \mathbb{R}^{M_k \times 1}$ denotes the measurement vector acquired at node $k$, $D_k \in \mathbb{R}^{M_k \times N}$ is the sensing matrix of the node $k$, $\theta_k \in \mathbb{R}^{N \times 1}$ is the $k'$th sparse signal, $n_k \in \mathbb{R}^{M_k \times 1}$ is the measurement noise for $k'$th signal, and $K$ is the total number of nodes. In the JSM-1 DCS model, the sparse signal $\theta_k$ can be represented as

$$\theta_k = w_c \odot z_c + w_k \odot z_k \qquad k = 1, ..., K \tag{10}$$

where $w_c \in \mathbb{R}^{N \times 1}$ is the common component, $z_c \in \{0,1\}^{N \times 1}$ is the support vector of $w_c$, $w_k \in \mathbb{R}^{N \times 1}$ is the innovation component of the $\theta_k$, which is specific to the $k'$th signal, $z_k \in \{0,1\}^{N \times 1}$ is the support vector of $w_k$, and $\odot$ denotes the Hadamard product.

In this paper, we utilize the BKF prior for the common and innovation components. Also, we assume an accurate prior for the elements of common and innovation supports, $z_c$ and $z_k$, which incorporates the intra- and inter-scale dependencies between the locations of the zero/ nonzero coefficients. Hence, the priors are given as

$$p(w_k) = BKF(a_k, b_k) \tag{11}$$

$$p(w_c) = BKF(a_c, b_c) \tag{12}$$

$$p(\lambda_{k,s}) = Gamma(\alpha_{k,s}, \beta_{k,s}) \tag{13}$$

$$p(\lambda_{c,s}) = Gamma(\alpha_{c,s}, \beta_{c,s}) \tag{14}$$

$$p(z_{k,s,i}) = Bernoulli(\pi_{k,s,i}) \tag{15}$$

$$p(z_{c,s,i}) = Bernoulli(\pi_{c,s,i}) \tag{16}$$

$$p(\alpha_n) = Gamma(c, d) \tag{17}$$

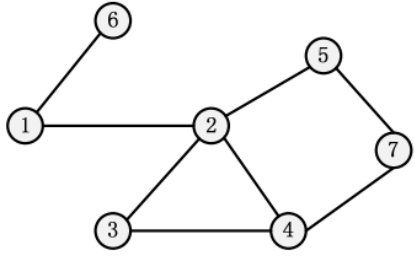

Fig. 2 A typical network structure with 7 nodes.

$$\pi_{k,s,i} = \begin{cases} \pi_{k,s} & s = 0,1 \\ \pi_{k,s00} & 2 \le s \le L, \ z_{pa(k,s,i)}=0, \ z_{\mathcal{N}(k,s,i)}=0 \\ \pi_{k,s01} & 2 \le s \le L, \ z_{pa(k,s,i)}=0, \ z_{\mathcal{N}(k,s,i)}=1 \\ \pi_{k,s10} & 2 \le s \le L, \ z_{pa(k,s,i)}=1, \ z_{\mathcal{N}(k,s,i)}=0 \\ \pi_{k,s11} & 2 \le s \le L, \ z_{pa(k,s,i)}=1, \ z_{\mathcal{N}(k,s,i)}=1 \end{cases} \quad (18)$$

$$\pi_{c,s,i} = \begin{cases} \pi_{c,s} & s = 0,1 \\ \pi_{c,s00} & 2 \le s \le L, \ z_{pa(c,s,i)}=0, \ z_{\mathcal{N}(c,s,i)}=0 \\ \pi_{c,s01} & 2 \le s \le L, \ z_{pa(c,s,i)}=0, \ z_{\mathcal{N}(c,s,i)}=1 \\ \pi_{c,s10} & 2 \le s \le L, \ z_{pa(c,s,i)}=1, \ z_{\mathcal{N}(c,s,i)}=0 \\ \pi_{c,s11} & 2 \le s \le L, \ z_{pa(c,s,i)}=1, \ z_{\mathcal{N}(c,s,i)}=1 \end{cases} \quad (19)$$

$$p(\pi_{k,s}) = Beta(e_{k,s}, f_{k,s}) \qquad s = 0,1 \tag{20.a}$$

$$p(\pi_{k,s00}) = Beta(e_{k,s00}, f_{k,s00}) \qquad s = 2, \dots, L \tag{20.b}$$

$$p(\pi_{k,s01}) = Beta(e_{k,s01}, f_{k,s01}) \qquad s = 2, \dots, L \tag{20.c}$$

$$p(\pi_{k,s10}) = Beta(e_{k,s10}, f_{k,s10}) \qquad s = 2, \dots, L \tag{20.d}$$

$$p(\pi_{k,s11}) = Beta(e_{k,s11}, f_{k,s11}) \qquad s = 2, \dots, L \tag{20.e}$$

$$p(\pi_{c,s}) = Beta(e_{c,s}, f_{c,s}) \qquad s = 0,1 \tag{21.a}$$

$$p(\pi_{c,s00}) = Beta(e_{c,s00}, f_{c,s00}) \qquad s = 2, \dots, L \tag{21.b}$$

$$p(\pi_{c,s01}) = Beta(e_{c,s01}, f_{c,s01}) \qquad s = 2, \dots, L \tag{21.c}$$

$$p(\pi_{c,s10}) = Beta(e_{c,s10}, f_{c,s10}) \qquad s = 2, \dots, L \tag{21.d}$$

$$p(\pi_{c,s11}) = Beta(e_{c,s11}, f_{c,s11}) \qquad s = 2, \dots, L \tag{21.e}$$

where $\lambda_{k,s}$ and $\lambda_{c,s}$ represent the variance of the innovation components and common components at scale $s$, respectively, i.e. we have adopted the same variance for the innovation/ common coefficients within the same scale, $(k,s,i)$ and $(c,s,i)$ is the $i'$th element index at scale $s$, which belongs to the $k'$th innovation component and the common component, respectively, $\pi_{k,s,i}$ and $\pi_{c,s,i}$ denote the mixing weights adopting Beta priors with the specified hyperparameters, $z_{pa(k,s,i)}$ and $z_{pa(c,s,i)}$ imply the inter-scale dependencies and denote the support of the parent coefficient of $w_{(k,s,i)}$ and $w_{(c,s,i)}$, respectively, $z_{\mathcal{N}(k,s,i)}$ and $z_{\mathcal{N}(c,s,i)}$ imply the intra-scale dependencies and represent the support of the second-order neighborhood for the $i'$th coefficient at scale $s$, and $L$ is the total number of wavelet decomposition levels. Moreover, $z_{pa(.,s,i)} = 0$ represents a zero-valued parent for the $i'$th coefficient, and $z_{pa(.,s,i)} = 1$ represents a nonzero parent for the $i'$th coefficient. Also, $z_{\mathcal{N}(.,s,i)} = 0$ denotes that the number of nonzero coefficients in the neighborhood of the $i'$th coefficient is smaller than the threshold, and $z_{\mathcal{N}(.,s,i)} = 1$ denotes that the number of nonzero coefficients in the neighborhood of the $i'$th coefficient is bigger than the threshold.

## 4. Variational Bayes Inference

We implement the posterior derivation based on the variational approximation in the Bayesian formulation of the proposed JSM-1 DCS, where the true posterior distribution $p(.)$ is estimated in a factorized form, say $q(.)$, and the Kullback–Leibler divergence between the posterior distribution and its approximation is minimized. Let $Y = \{y_1, \dots, y_K\}$, $W = \{w_c, w_1, \dots, w_K\}$, $Z = \{z_c, z_1, \dots, z_K\}$ and $\theta = \{\Gamma_c, \Gamma_1, \dots, \Gamma_K, \pi_c, \pi_1, \dots, \pi_K, \alpha_n\}$, where $\Gamma_k$ ($k = 1, \dots, K$) and $\Gamma_c$ are $N \times N$ diagonal matrices whose elements are the variances $\lambda_{k,s}$ and $\lambda_{c,s}$, respectively. For the proposed JSM-1 DCS setting, we assume that the approximate posteriors $q(W)$ and $q(Z)$ can be factorized as

$$q(W) = q(w_c)q(w_1) \dots q(w_K) \tag{22}$$

$$q(Z) = q(z_c)q(z_1) \dots q(z_K) \tag{23}$$

According to the prior distributions presented in the section 3, it can be proved that the common component and the innovation components have Gaussian distributions. The updated formula for the posterior distributions $q(w_c)$ and $q(w_k)$ at each iteration can be expressed as

$$q(w_c) \propto \exp(\mathbb{E}_{q(w_1),\dots,q(w_K)}[\ln p(y_1, \dots, y_K, w_c, w_1, \dots, w_K$$
$$, z_c, z_1, \dots, z_K; \Gamma_c, \Gamma_1, \dots, \Gamma_K, \pi_c, \pi_1, \dots, \pi_K, \alpha_n)])$$

$$\propto \exp(\mathbb{E}_{q(w_1)}[\ln p(y_1|w_c, w_1, z_c, z_1, \alpha_n)] + \dots + \mathbb{E}_{q(w_K)}[\ln p(y_K|w_c, w_K, z_c, z_1, \alpha_n)] + \ln p(w_c|\Gamma_c))$$

$$= \mathcal{N}(\mu_c, \Sigma_c) \tag{24}$$

$$q(w_k) \propto \exp(\mathbb{E}_{q(w_c),q(w_j),j \neq k}[\ln p(y_1, \dots, y_K, w_c, w_1, \dots,$$
$$w_K, z_c, z_1, \dots, z_K; \Gamma_c, \Gamma_1, \dots, \Gamma_K, \pi_c, \pi_1, \dots, \pi_K, \alpha_n)])$$

$$\propto \exp(\mathbb{E}_{q(w_c)}[\ln p(y_k|w_c, w_k, z_c, z_k, \alpha_n)] + \ln p(w_k|\Gamma_k))$$

$$= \mathcal{N}(\mu_k, \Sigma_k) \tag{25}$$

where $\mathbb{E}_{(.)}(.)$ represents the expectation with respect to the random variable in its subscript, and we have

$$\mu_c = \alpha_n \Sigma_c Z_c^T \left( \sum_{k=1}^{K} \{D_k^T (y_k - D_k Z_k \mu_k)\} \right) \tag{26.a}$$

$$\Sigma_c = \left\{ \Gamma_c + \alpha_n Z_c^T \left( \sum_{k=1}^{K} D_k^T D_k \right) Z_c \right\}^{-1} \tag{26.b}$$

$$\mu_k = \alpha_n \Sigma_k Z_k^T D_k^T (y_k - D_k Z_c \mu_c) \tag{27.a}$$

$$\Sigma_k = (\alpha_n Z_k^T D_k^T D_k Z_k + \Gamma_k)^{-1} \tag{27.b}$$

where $Z_{(.)} = diag(z_{(.)})$.

Now given $q(w_c)$ and $q(w_k)$, the variational distributions are obtained by applying similar techniques for the support vector as follows:

$$q(z_{k,s,i}) \propto p(z_{k,i}) \sqrt{\Sigma_{k,i}} \, exp\left(-\frac{1}{2} \frac{\mu_{k,i}^2}{\Sigma_{k,i}}\right) \tag{28}$$

$$q(z_{c,s,i}) \propto p(z_{c,i}) \sqrt{\Sigma_{c,i}} \, exp\left(-\frac{1}{2} \frac{\mu_{c,i}^2}{\Sigma_{c,i}}\right) \tag{29}$$

Also, the posteriors corresponding to the variances $\lambda_{k,s}$ and $\lambda_{c,s}$ has the Generalized Inverse Gaussian (GIG) distribution given by

$$q(\lambda_{k,s}) = GIG(\alpha'_{k,s}, \beta'_{k,s}, p_{k,s}) \tag{30}$$

$$q(\lambda_{c,s}) = GIG(\alpha'_{c,s}, \beta'_{c,s}, p_{c,s}) \tag{31}$$

where $\alpha'_{k,s} = 2\beta_{k,s}$, $\beta'_{k,s} = \langle w_{k,i}^2 \rangle$, $p_{k,s} = \beta_{k,s} - \frac{1}{2}$, $\alpha'_{c,s} = 2\beta_{c,s}$, $\beta'_{c,s} = \langle w_{c,i}^2 \rangle$, $p_{c,s} = \beta_{c,s} - \frac{1}{2}$, and the angle bracket $\langle . \rangle$ represents the expected value of the variable within the brackets.

Because of the conjugacy property of the Gamma distribution and the Gaussian distribution, the posterior of the noise precision will take the form of Gamma distribution and is estimated based on the following

$$q(\alpha_n) = Gamma(c', d') \tag{32}$$

where

$$c' = c + \frac{KM}{2} \tag{33.a}$$

$$d' = d + \frac{1}{2} \sum_{k=1}^{K} \|y_k - D_k \theta_k\|_2^2 \tag{33.b}$$

According to (18) - (21), the posterior distributions corresponding to the $\pi_{k,s,i}$ and $\pi_{c,s,i}$ are obtained, which exploits both the intra- and inter-scale correlations

$$q(\pi_k) = \prod_{s=0}^{1} Beta(e'_{k,s}, f'_{k,s}) \times$$
$$\prod_{s=2}^{L} Beta(e'_{k,s00}, f'_{k,s00}) Beta(e'_{k,s01}, f'_{k,s01}) Beta(e'_{k,s10}, f'_{k,s10}) Beta(e'_{k,s11}, f'_{k,s11}) \tag{34}$$

$$q(\pi_c) = \prod_{s=0}^{1} Beta(e'_{c,s}, f'_{c,s}) \times$$
$$\prod_{s=2}^{L} Beta(e'_{c,s00}, f'_{c,s00}) Beta(e'_{c,s01}, f'_{c,s01}) Beta(e'_{c,s10}, f'_{c,s10}) Beta(e'_{c,s11}, f'_{c,s11}) \quad (35)$$

where the hyperparameters of the above posteriors are given by

$$e'_{k,s} = e_{k,s} + \sum_i z_{i,k,s} \tag{36.a}$$

$$f'_{k,s} = f_{k,s} + M_s - \sum_i (1 - z_{i,k,s}) \tag{36.b}$$

$$e'_{k,s00} = e_{k,s00} + \sum_i (1 - z_{pa(k,s,i)})(1 - z_{\mathcal{N}(k,s,i)}) z_{i,k,s} \tag{36.c}$$

$$f'_{k,s00} = f_{k,s00} + \sum_i (1 - z_{pa(k,s,i)})(1 - z_{\mathcal{N}(k,s,i)})(1 - z_{i,k,s}) \tag{36.d}$$

$$e'_{k,s01} = e_{k,s01} + \sum_i (1 - z_{pa(k,s,i)}) z_{\mathcal{N}(k,s,i)} z_{i,k,s} \tag{36.e}$$

$$f'_{k,s01} = f_{k,s01} + \sum_i (1 - z_{pa(k,s,i)}) z_{\mathcal{N}(k,s,i)} (1 - z_{i,k,s}) \tag{36.f}$$

$$e'_{k,s10} = e_{k,s10} + \sum_i z_{pa(k,s,i)} (1 - z_{\mathcal{N}(k,s,i)}) z_{i,k,s} \tag{36.g}$$

$$f'_{k,s10} = f_{k,s10} + \sum_i z_{pa(k,s,i)} (1 - z_{\mathcal{N}(k,s,i)}) (1 - z_{i,k,s}) \tag{36.h}$$

$$e'_{k,s11} = e_{k,s11} + \sum_i z_{pa(k,s,i)} z_{\mathcal{N}(k,s,i)} z_{i,k,s} \tag{36.i}$$

$$e'_{k,s11} = e_{k,s11} + \sum_i z_{pa(k,s,i)} z_{\mathcal{N}(k,s,i)} (1 - z_{i,k,s}) \tag{36.j}$$

$$e'_{c,s} = e_{c,s} + \sum_i z_{i,c,s} \tag{37.a}$$

$$f'_{c,s} = f_{c,s} + M_s - \sum_i (1 - z_{i,c,s}) \tag{37.b}$$

$$e'_{c,s00} = e_{c,s00} + \sum_i (1 - z_{pa(c,s,i)})(1 - z_{\mathcal{N}(c,s,i)}) z_{i,c,s} \tag{37.c}$$

$$f'_{c,s00} = f_{c,s00} + \sum_i (1 - z_{pa(c,s,i)})(1 - z_{\mathcal{N}(c,s,i)})(1 - z_{i,c,s}) \tag{37.d}$$

$$e'_{c,s01} = e_{c,s01} + \sum_i (1 - z_{pa(c,s,i)}) z_{\mathcal{N}(c,s,i)} z_{i,c,s} \tag{37.e}$$

$$f'_{c,s01} = f_{c,s01} + \sum_i (1 - z_{pa(c,s,i)}) z_{\mathcal{N}(c,s,i)} (1 - z_{i,c,s}) \tag{37.f}$$

$$e'_{c,s10} = e_{c,s10} + \sum_i z_{pa(c,s,i)} (1 - z_{\mathcal{N}(c,s,i)}) z_{i,c,s} \tag{37.g}$$

$$f'_{c,s10} = f_{c,s10} + \sum_i z_{pa(c,s,i)} (1 - z_{\mathcal{N}(c,s,i)}) (1 - z_{i,c,s}) \tag{37.h}$$

$$e'_{c,s11} = e_{c,s11} + \sum_i z_{pa(c,s,i)} z_{\mathcal{N}(c,s,i)} z_{i,c,s} \tag{37.i}$$

$$e'_{c,s11} = e_{c,s11} + \sum_i z_{pa(c,s,i)} z_{\mathcal{N}(c,s,i)} (1 - z_{i,c,s}) \tag{37.j}$$

Since the VB is ensured to converge, the variational approximation procedure iteratively updates (24) to (37) until convergence occurs. Finally, the reconstructed sparse signal can be obtained as

$$\theta_k = \mu_c \odot z_c + \mu_k \odot z_k \quad k = 1, \dots, K \tag{38}$$

The WBDCS-BKF algorithm explained above is a centralized approach, which converges to the local maximum of $\log p(Y)$ by repeatedly iterating the VB updating procedure, and thus simplifies the design of our proposed decentralized algorithm that presented in the next section.

## 5. Wavelet-based Decentralized Bayesian Algorithm for DCS

In this section, we propose a decentralized version of the centralized JSM-1 WBDCS-BKF algorithm developed in the previous section. Our objective is to design an algorithm to solve the joint reconstruction problem in the absence of a fusion center, where each node communicates non-sensitive information with its first-order neighbors only. Thus, the main issue in the decentralized approach is the parallel implementation of the VB inference procedure, so that the reconstruction of the innovation components and common component is decoupled. While in the centralized approach, the update equations presented by (26.a), (26.b), and (33.b) is computed at a fusion center, which obtains all the nodes information and computes the posteriors.

The equations (26.a), (26.b), and (33.b) of the centralized VB inference can be recast as

$$\frac{1}{K\alpha_n} Z_c^{T^{-1}} (\Sigma_c^{-1} - \Gamma_c) Z_c^{-1} = \frac{1}{K} \sum_{k=1}^{K} D_k^T D_k \tag{39}$$

$$\frac{1}{K\alpha_n} \Sigma_c^{-1} Z_c^{T^{-1}} \mu_c = \frac{1}{K} \sum_{k=1}^{K} \{D_k^T (y_k - D_k Z_k \mu_k)\} \tag{40}$$

$$\frac{2}{K}(d' - d) = \frac{1}{K} \sum_{k=1}^{K} \|y_k - D_k \theta_k\|_2^2 \tag{41}$$

where the above equations represent the averages of $D_k^T D_k$, $D_k^T(y_k - D_k Z_k \mu_k)$ and $\|y_k - D_k \theta_k\|_2^2$ respectively, and can be solved by reformulating as the following optimization problem

$$\min_{A} \sum_{k=1}^{K} \|A - D_k^T D_k\|_F^2 \tag{42}$$

$$\min_{\gamma} \sum_{k=1}^{K} \|\gamma - D_k^T (y_k - D_k Z_k \mu_k)\|_2^2 \tag{43}$$

$$\min_{\lambda} \sum_{k=1}^{K} \|\lambda - \|y_k - D_k \theta_k\|_2^2 \|_2^2 \tag{44}$$

To solve this minimization problems, we reformulate them to have the form of a separable objective function with a set of respective constraints

$$\min_{A^1,\ldots,A^K} \sum_{k=1}^{K} \|A^k - D_k^T D_k\|_F^2 \quad s.t. \quad A^k = A^{j_k} \quad \forall j_k \in \mathcal{N}_k, \ \forall k \in \{1,\ldots,K\} \tag{45}$$

$$\min_{\gamma^1,\ldots,\gamma^K} \sum_{k=1}^{K} \|\gamma^k - D_k^T (y_k - D_k Z_k \mu_k)\|_2^2 \quad s.t. \quad \gamma^k = \gamma^{j_k} \quad \forall j_k \in \mathcal{N}_k, \ \forall k \in \{1,\ldots,K\} \tag{46}$$

$$\min_{\lambda^1,\ldots,\lambda^K} \sum_{k=1}^{K} \|\lambda^k - \|y_k - D_k \theta_k\|_2^2 \|_2^2 \quad s.t. \quad \lambda^k = \lambda^{j_k} \quad \forall j_k \in \mathcal{N}_k, \ \forall k \in \{1,\ldots,K\} \tag{47}$$

where $A^k$, $\gamma^k$ and $\lambda^k$ are the local estimates of $A = \frac{1}{K}\sum_{k=1}^{K} D_k^T D_k$, $\gamma = \frac{1}{K}\sum_{k=1}^{K} D_k^T(y_k - D_k Z_k \mu_k)$ and $\frac{1}{K}\sum_{k=1}^{K} \|y_k - D_k \theta_k\|_2^2$ computed at node $k$, respectively, and $\mathcal{N}_k$ is the set of neighbors of node $k$. We employ the alternating direction method of multipliers (ADMM) to solve the above constrained minimization problems. According to [37], the estimation error can be minimized by iteratively updating local estimates as follows

$$\left(B_A^k\right)^{new} = \left(B_A^k\right)^{old} + \rho \sum_{j_k \in \mathcal{N}_k} \left(\left(A^k\right)^{old} - \left(A^{j_k}\right)^{old}\right) \tag{48.a}$$

$$\left(A^k\right)^{new} = \frac{1}{2+2\rho|\mathcal{N}_k|}\left(2D_k^T D_k - \left(B_A^k\right)^{new} + \rho \sum_{j_k \in \mathcal{N}_k}\left(\left(A^k\right)^{old} + \left(A^{j_k}\right)^{old}\right)\right) \tag{48.b}$$

$$\left(B_\gamma^k\right)^{new} = \left(B_\gamma^k\right)^{old} + \rho \sum_{j_k \in \mathcal{N}_k}\left(\left(\gamma^k\right)^{old} - \left(\gamma^{j_k}\right)^{old}\right) \tag{49.a}$$

$$\left(\gamma^k\right)^{new} = \frac{1}{2+2\rho|\mathcal{N}_k|}\left(2D_k^T(y_k - D_k Z_k \mu_k) - \left(B_\gamma^k\right)^{new} + \rho \sum_{j_k \in \mathcal{N}_k}\left(\left(\gamma^k\right)^{old} + \left(\gamma^{j_k}\right)^{old}\right)\right) \tag{49.b}$$

$$\left(B_\lambda^k\right)^{new} = \left(B_\lambda^k\right)^{old} + \rho \sum_{j_k \in \mathcal{N}_k}\left(\left(\lambda^k\right)^{old} - \left(\lambda^{j_k}\right)^{old}\right) \tag{50.a}$$

$$\left(\lambda^k\right)^{new} = \frac{1}{2+2\rho|\mathcal{N}_k|}\left(2\|y_k - D_k\theta_k\|_2^2 - \left(B_\lambda^k\right)^{new} + \rho \sum_{j_k \in \mathcal{N}_k}\left(\left(\lambda^k\right)^{old} + \left(\lambda^{j_k}\right)^{old}\right)\right) \tag{50.b}$$

for $k = 1, \ldots, K$, where $\rho$ is a positive coefficient. Note that nodes can compute (48) in parallel and independent of VB iterations, while (49) and (50) should be execute in each VB iteration when $\mu_k$ and $Z_k$ are updated.

Substituting $A^k$, $\gamma^k$ and $\lambda^k$ into (26.a), (26.b), and (33.b), respectively, we have

$$\Sigma_c = \{\Gamma_c + K\alpha_n Z_c^T A Z_c\}^{-1} \tag{51}$$

$$\mu_c = \alpha_n \Sigma_c Z_c^T \gamma \tag{52}$$

$$d' = d + \frac{1}{2}\lambda \tag{53}$$

The summary of the proposed decentralized algorithm for DCS signal reconstruction via WBDCS-BKF algorithm is given in Algorithm 1. As it is evident from this pseudo-code, the VB procedure proceeds by iteratively updating parameters and hidden variables until a fixed point is reached or the number of iterations exceeds a predefined number.

## 6. Computational complexity

In this subsection, we analyze the computational complexity of the proposed algorithm and compare it with four CS recovery algorithms. The computational complexity of the BKF-based centralized algorithm proposed in this paper is dominated by the expectation step (E-step), i.e., (26) and (27). Let, similar to the previous sections, $K$ be the number of network nodes. Thus, (26) involves $\mathcal{O}(M'N^2 + N^3)$ multiplications per iteration, where $M' = \sum_{k=1}^{K} M_k$; and (27) involves $\mathcal{O}(M_k N^2 + N^3)$ multiplications per iteration for the $k$−th node. Hence, the overall asymptotic computational complexity is of the order of $\mathcal{O}(M'N^2 + KN^3)$ per iteration. On the other hand, the only difference between the centralized and decentralized versions of the proposed BKF-based algorithm is in the updating formulas of the common components. Thus, the computational complexity of the proposed decentralized algorithm is dominated by (51), (52), and (27), which results a complexity of the order of $\mathcal{O}(M'N^2 + KN^3)$ per iteration. Moreover, the computational complexity of the algorithm presented in [15] is $\mathcal{O}(M'N^2 + KN^3)$ per iteration. In comparison, the overall computational complexity of the TSWCS algorithm [36], SA-SBL algorithm [38], and FH-BCS algorithm [39], which are not distributed methods, is of the order of $\mathcal{O}((NK)^3)$ per iteration, which demonstrates that jointly reconstruction of multiple correlated sparse signals reduces the computational complexity of the algorithm.

---

Algorithm 1: WBDCS-BKF Algorithm

---

**Input:** A set of measurements $\{y_k\}_{k=1}^K$, a set of sensing matrices $\{A_k\}_{k=1}^K$, and the maximum number of iterations max_iter.

**Process:**

1) Initialize $\Sigma_c$ and $\{\Sigma_k\}_{k=1}^K$ by identity matrices $I_{N \times N}$ and $\mu_c, z_c, \{\mu_k, z_k\}_{k=1}^K$ by $\mathbf{0}_{N \times 1}$;

2) Initialize $\{B_A^k\}_{k=1}^K$ and $\{A^k\}_{k=1}^K$ by $\mathbf{0}_{N \times N}$ and iteratively compute (48) in parallel at each node until a predefined halting criterion is satisfied;

3) Initialize $\{B_\gamma^k\}_{k=1}^K$ and $\{\gamma^k\}_{k=1}^K$ by $\mathbf{0}_{N \times 1}$ and iteratively compute (49) in parallel at each node until a predefined halting criterion is satisfied;

4) Compute $\Sigma_c = \{\Gamma_c + K\alpha_n Z_c^T A Z_c\}^{-1}$, $\mu_c = \alpha_n \Sigma_c Z_c^T \gamma$, (29) and (35) at each node;

5) Compute (27), (28) and (34) at each node;

6) Compute (30) and (31) at each node;

7) Initialize $\{B_\lambda^k\}_{k=1}^K$ and $\{\lambda^k\}_{k=1}^K$ by 0 and iteratively compute (50) in parallel at each node until a predefined halting criterion is satisfied;

8) Compute $d' = d + \frac{1}{2}\lambda$ and (32);

9) If stopping condition is satisfied, compute $\theta_k = \mu_c \odot z_c + \mu_k \odot z_k$ for each node, else go to step (2).

**Output:** $\{\theta_k\}_{k=1}^K$

---

# 6. Simulation results

In this section, we investigate the performance of the proposed WBDCS-BKF algorithm in both the unstructured and structured cases and compare the results with that of state-of-the-art algorithms in terms of the normalized mean square error (NMSE), peak signal to noise ratio (PSNR), convergence rate, and running time. The NMSE is defined as $NMSE = \frac{1}{K}\sum_{k=1}^{K} \frac{\|x_k - \widehat{x_k}\|_2^2}{\|x_k\|_2^2}$ where $K$ is the total number of nodes, $x_k$ and $\widehat{x_k}$ denote the $k$'th original sparse signal and the reconstructed sparse signal, respectively.

In the following experiments, we use the DWT transform as the sparse domain, and the components of all the measurement matrices $D_k$ are randomly drawn from a zero-mean Gaussian distribution with variance $\frac{1}{M}$.

## 6.1 Analysis of performance in unstructured case

In order to investigate the potential of the BKF prior in modeling the DWT coefficients, we consider the unstructured case, in which the statistical structure associated with the DWT coefficients is not taken into account, and the only assumption is the intra- and inter-signal dependencies among the multiple sparse signals. We compare the performance of the proposed BKF-based DCS algorithm with that of the Gaussian-based Bayesian DCS proposed by [15]. Both the algorithms assume JSM-1 model for the signals. In [15], the discrete cosine transform (DCT) is used as the sparsifying domain, and the random partial DCT matrices are used as the measurement matrices. For the sake of fair comparison, we used the DWT transform and Gaussian measurement matrices in all simulations.

In this scenario, we consider a set of $K = 10$ one-dimensional EEG signals of length $N = 512$, which are downloaded from [40]. Fig. 3 illustrates the algorithm performance in terms of the NMSE as a function of the sampling rate $(M/N)$ at each node. To evaluate each point in the curves, 100 trials are averaged. As can be seen from Fig. 3, for the proposed BKF-based DCS algorithm, the reconstruction error is lower than the Gaussian-based DCS algorithm. This is due to the assumption of an accurate prior distribution, which has a sharp peak at zero and heavier tail than the Gaussian pdf and, thus, can demonstrate the intra-scale dependencies of wavelet coefficients more efficient than the Gaussian pdf.

Fig. 4 represents the convergence rates of the proposed BKF-based DCS and the Gaussian-based DCS algorithm for the EEG signals mentioned above. It is observed that the proposed algorithm exhibits reasonably fast convergence. In particular, after 40 loop iterations, the proposed algorithm becomes flat and stable, which demonstrates the convergence of the algorithm. While the convergence of the Gaussian-based DCS algorithm occurs after 70 iterations.

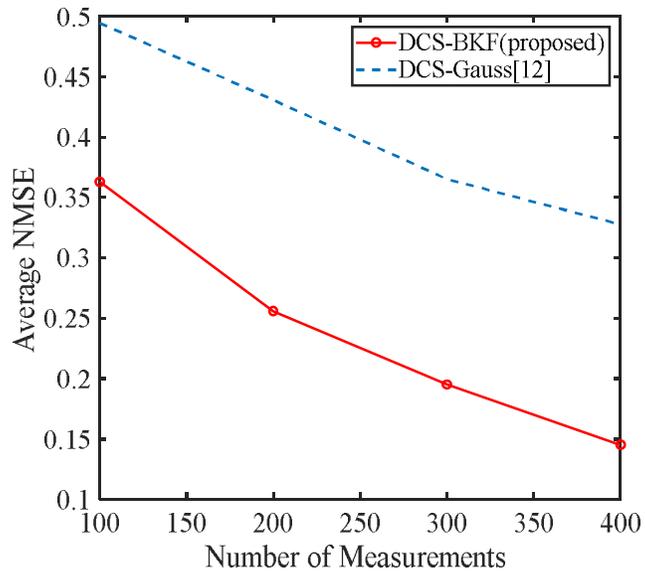

Fig. 3 Comparison of the normalized mean squared error for $K = 10$ EEG signals.

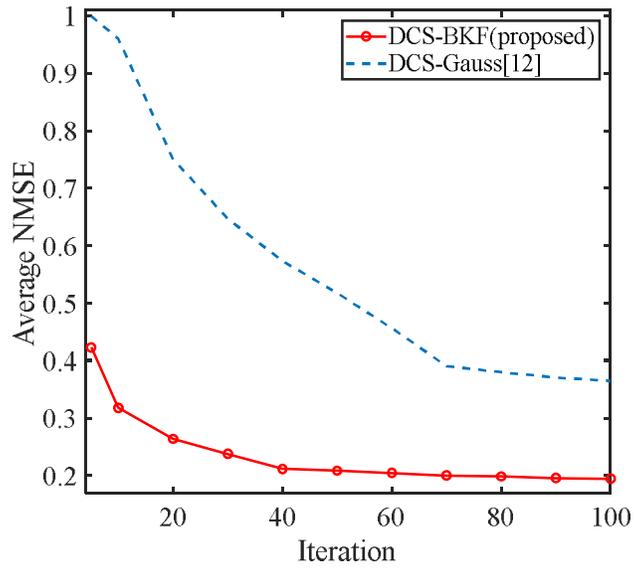

Fig. 4 Convergence rate comparison for $K = 10$ EEG signals.

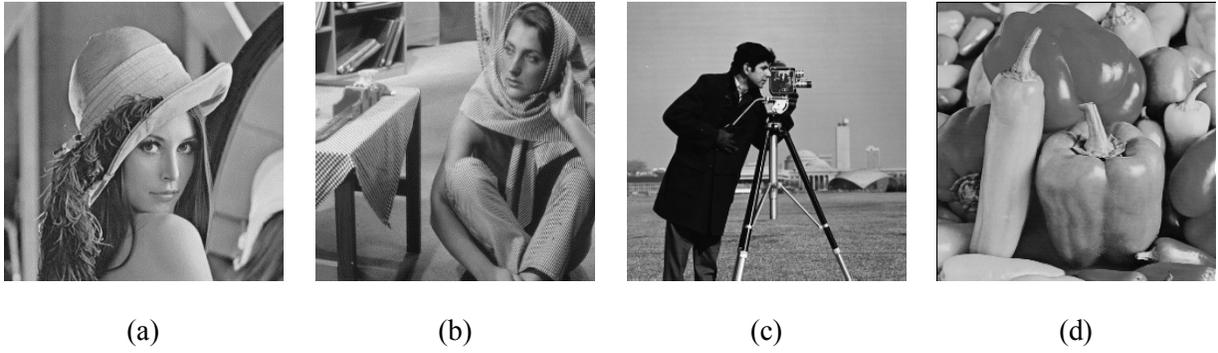

Fig. 5 The test images used in the experiments: (a) "Lena", (b) "Barbara", (c) "Cameraman", and (d) "Pepper".

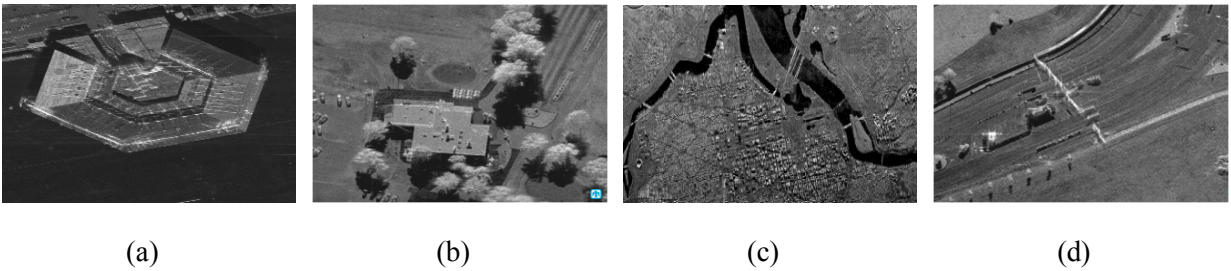

Fig. 6 The SAR images used in the experiments [42].

**6.2 Analysis of performance in structured case**

In the structured case, the statistical dependencies between the locations of atoms are taken into account in the sparse wavelet coefficients. To evaluate the proposed WBDCS-BKF algorithm, we compare the performance of three different algorithms recently proposed for the DCS recovery problem: the Bayesian DCS algorithm [15], the backtracking-based adaptive orthogonal matching pursuit for block DCS (DCSBBAOMP) algorithm [41], Tree-Structured Wavelet-based CS using MCMC (TSWCS) [36], SA-SBL [38] (which is an algorithm based on structure-aware sparse Bayesian learning (SA-SBL) and employs approximate message passing to accelerate the learning process), FH-BCS [39] (which is a structure-aware BCS algorithm to estimate the signal spectrum.).The distributed algorithm presented in [15] is a Bayesian algorithm, which follows JSM-1 model and adopts Gaussian distribution for the common and the innovation components, and then, employs VB inference for estimating the posterior probabilities of multiple sparse signals. The DCSBBAOMP algorithm is an OMP-based method, which jointly reconstructs block-sparse signals in an iterative manner. Each iteration of this algorithm consists of forward selection and backward removal stages.

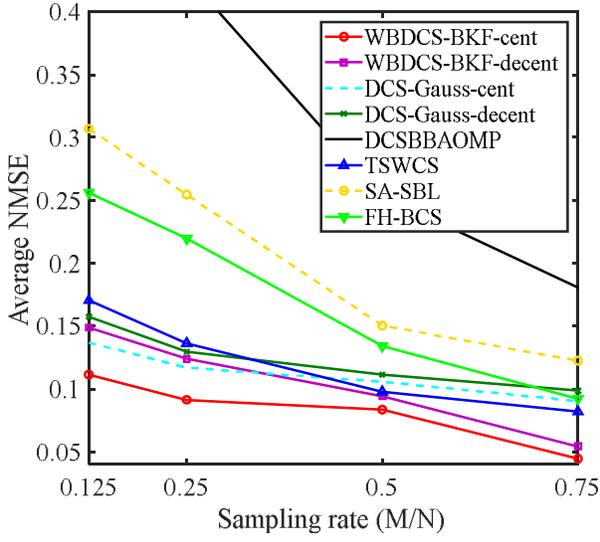 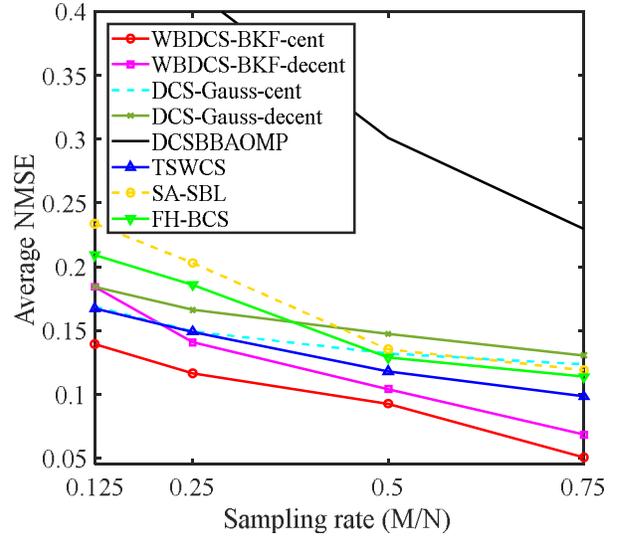

(a)  (b)

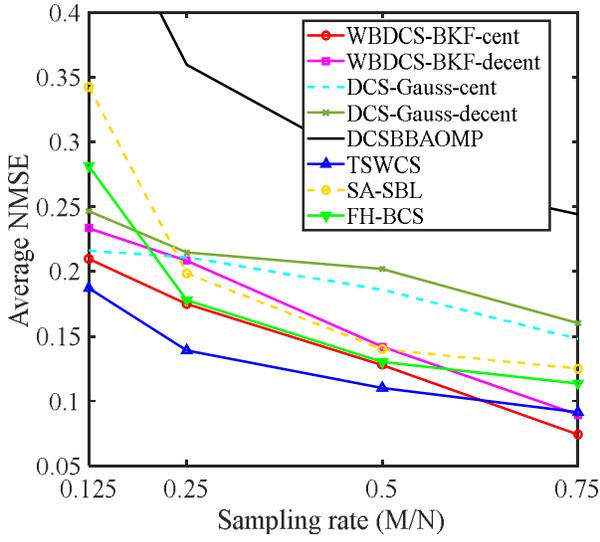 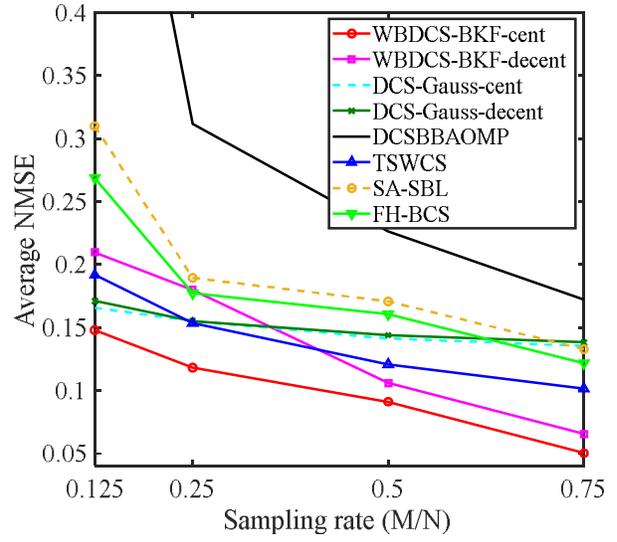

(c)  (d)

Fig. 7: Normalized mean square error as a function sampling rate ($M/N$) for the optical images presented in Fig. 5: (a) "Lena", (b) "Barbara", (c) "Cameraman", and (d) "Pepper".

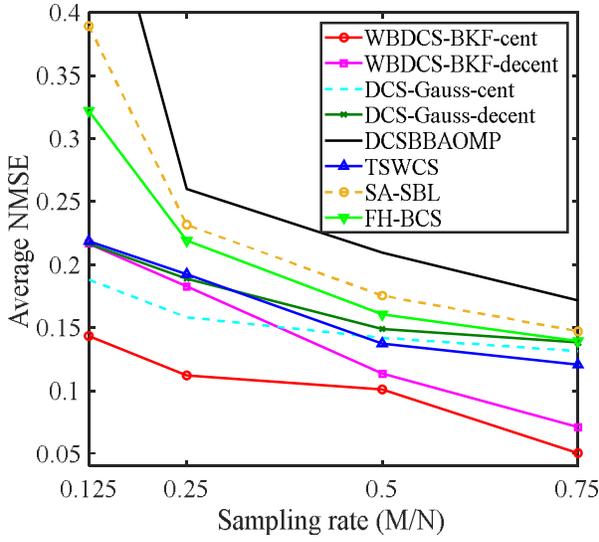 (a)

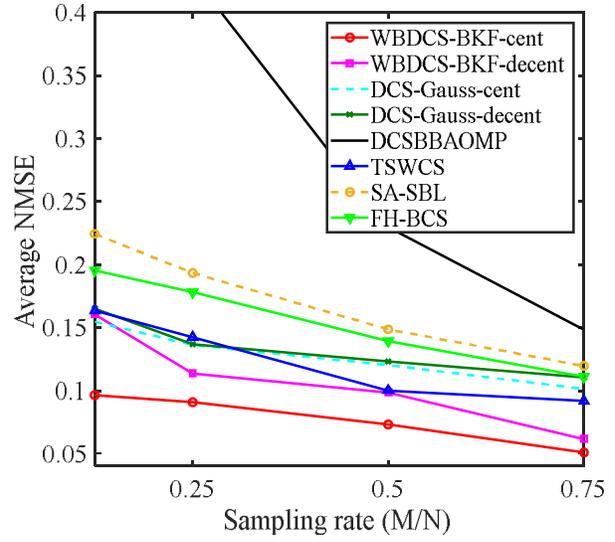 (b)

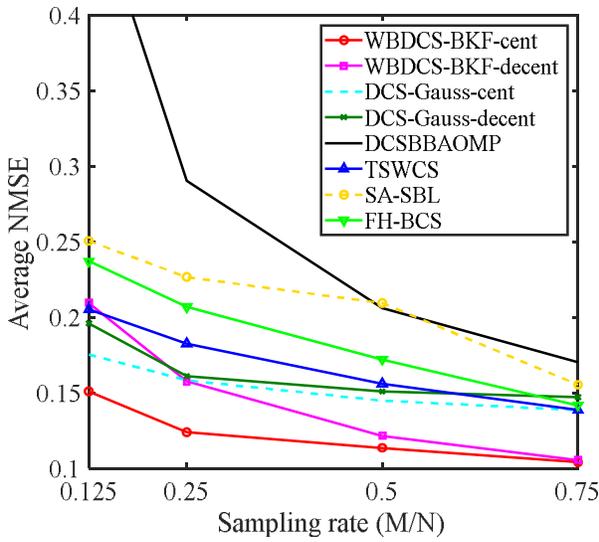 (c)

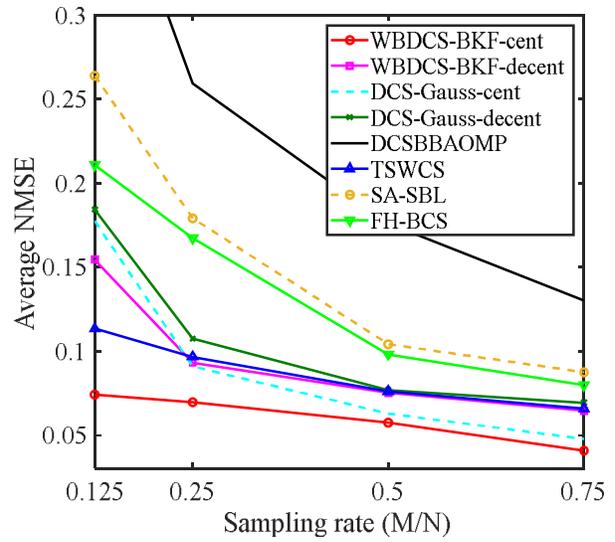 (d)

Fig. 8: Normalized mean square error as a function of sampling rate ($M/N$) for the SAR images presented in Fig. 6.

To investigate the performance of the proposed WBDCS-BKF algorithm, we consider four $128 \times 128$ optical images and four $128 \times 128$ SAR images. The optical images shown in fig. 5 include: 'Lena', 'Barbara', 'Cameraman', and 'Pepper' images. SAR images are downloaded from Sandia National Laboratories in U.S. [42] and shown in fig. 6. Similar to the process of jointly sparse signals generation described in the literature [41], we first cut each image into $32 \times 32$ blocks, such that the total number of 16 blocks are obtained and these blocks are jointly sparse in the wavelet domain. Then, we perform the CS measurement procedure for each of the blocks, and finally, we exploit the joint reconstruction algorithm. The advantage of such processing is that the reconstruction procedure divides into some smaller and simpler processes.

In addition, in the simulation of decentralized algorithms, a $P$-connected Harary graph is applied to form the neighborhood relationship [15], where each node is only allowed to exchange information with its $P$ neighbors. In the experiments, we consider $P = 5$.

Figs. 7 and 8 plots NMSE of the considered algorithms against the sampling rate for optical and SAR images, respectively. To calculate each point of these curves, 50 trials are averaged. Due to the efficiently employment of both intra- and inter-scale dependencies in the proposed WBDCS-BKF reconstruction algorithm, the reconstruction error decreases significantly in both the centralized and decentralized cases. Moreover, comparing the curves presented in this figures, it can be observed that in most cases, the NMSE of the proposed algorithm is lower than that of the conventional CS algorithms, which asserts the efficiency of our algorithm which introduces a recovery process with reduced computational complexity and lower NMSE.

It can be observed from figs. 7 and 8 that the reconstruction error of the proposed decentralized approach is slightly more than that of the centralized approach. This is because of employment of VB procedure in the decentralized computations and decoupling the common component from innovation components, which prevents sharing sensitive information between nodes, and provides an approximation to the (26.a), (26.b), and (33.b) of the centralized VB inference. However, the centralized approach suffers from any danger at the FC, which results in the entire system failing.

The PSNR results of the competing algorithms against the sampling rate are presented in figs. 9 and 10 for optical and SAR images, respectively. It is observed that in both the centralized and decentralized approaches, the proposed algorithm has higher PSNR than that of the others.

The convergence rates of the proposed algorithms is illustrated in fig. 11 for the 'Pepper' image shown in fig. 5 and for $M = 400$, where we have compared the NMSE against the iteration number. It can be observed that as the iteration number increases, the NMSE of both the centralized and decentralized algorithms decrease monotonically. Moreover, the algorithms become flat and stable after around 40 iterations, which demonstrates the convergence of the proposed algorithm.

A running time comparison between the proposed BKF-based DCS algorithm and the other competing state-of-the-art methods in CS recovery is shown in fig. 12 for the 'Pepper' image shown in fig. 5. As it can be seen from fig. 12, running time of the proposed algorithm is slightly more than that of [15]. This is due to additional calculations belong to the HMT model employed in the proposed algorithm. In addition, as expected, the running time of the distributed algorithms is less than that of the non-distributed algorithms.

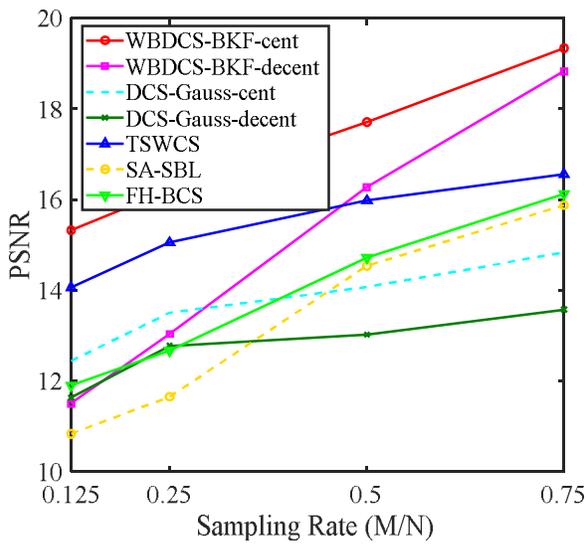
(a)

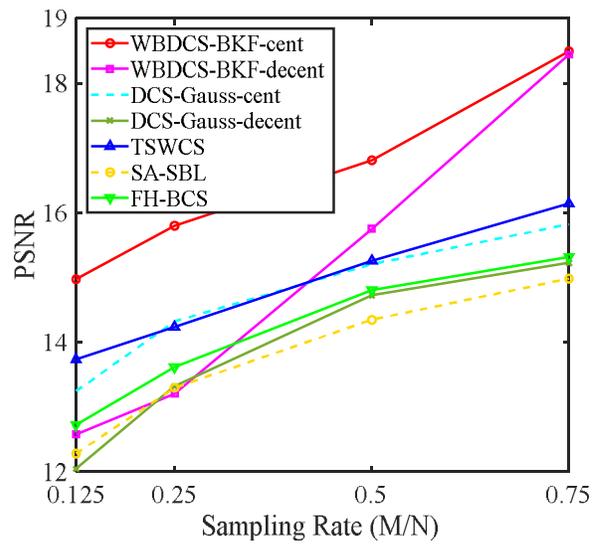
(b)

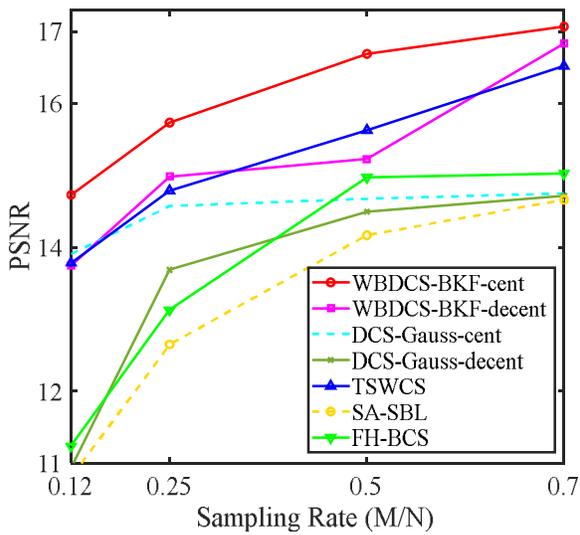
(c)

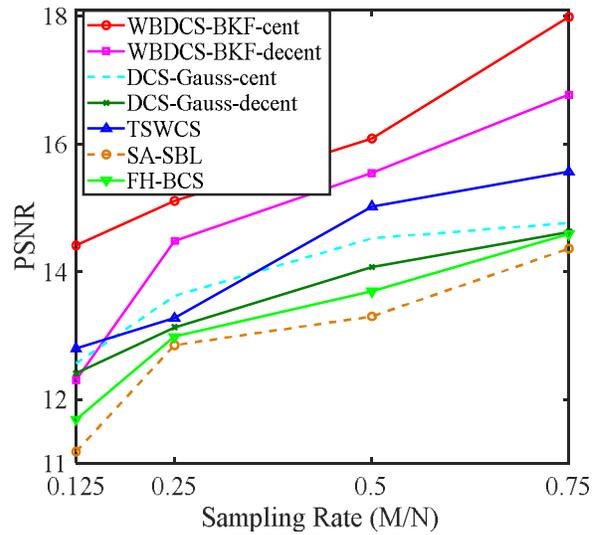
(d)

Fig. 9: PSNR comparison as a function sampling rate ($M/N$) for the optical images presented in Fig. 5: (a) "Lena", (b) "Barbara", (c) "Cameraman", and (d) "Pepper".

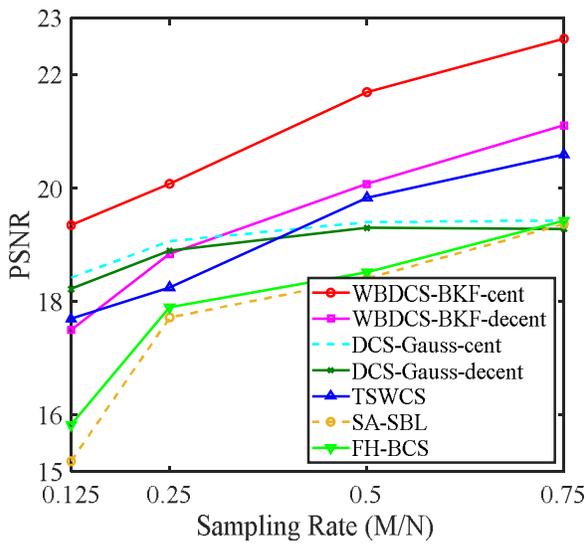
(a)

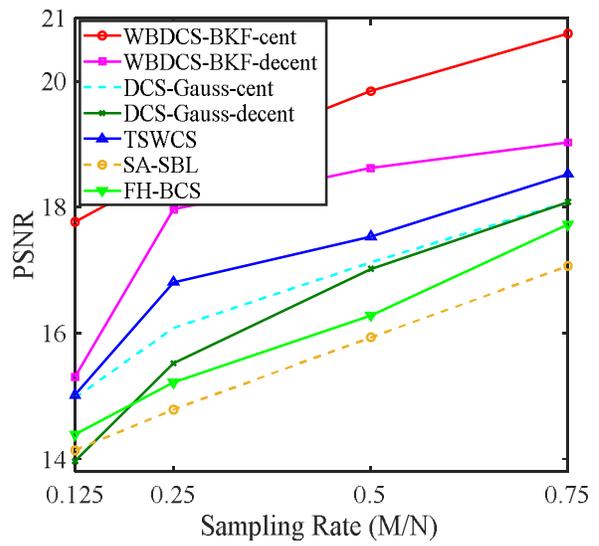
(b)

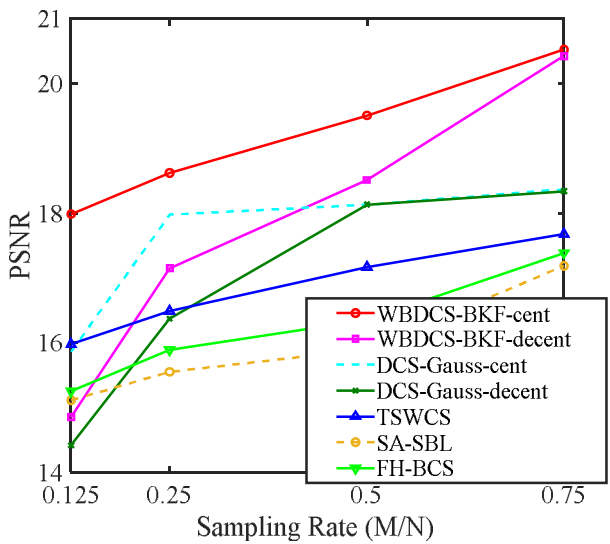
(c)

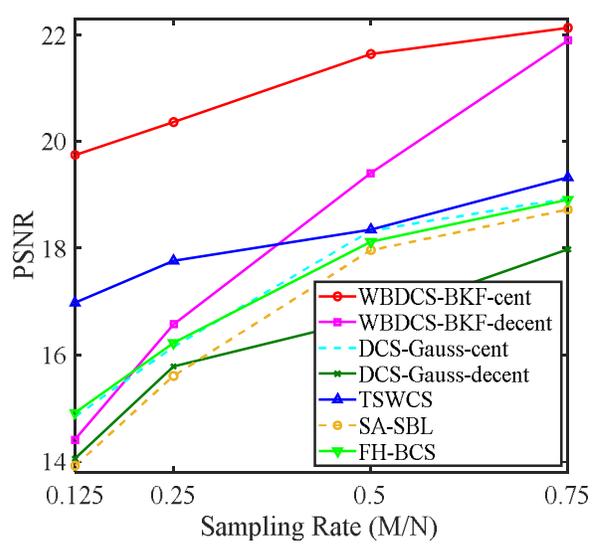
(d)

Fig. 10: PSNR comparison as a function of sampling rate ($M/N$) for the SAR images presented in Fig. 6.

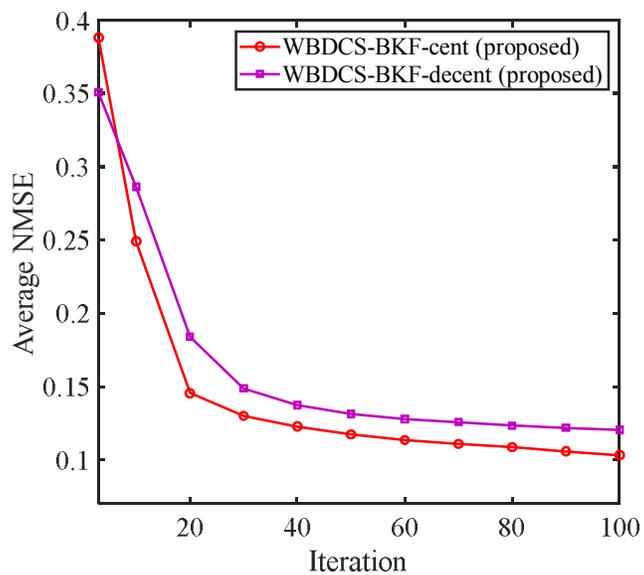

Fig. 11: Convergence rate of the proposed algorithm for the 'Pepper' image shown in fig. 5 and $N = 400$.

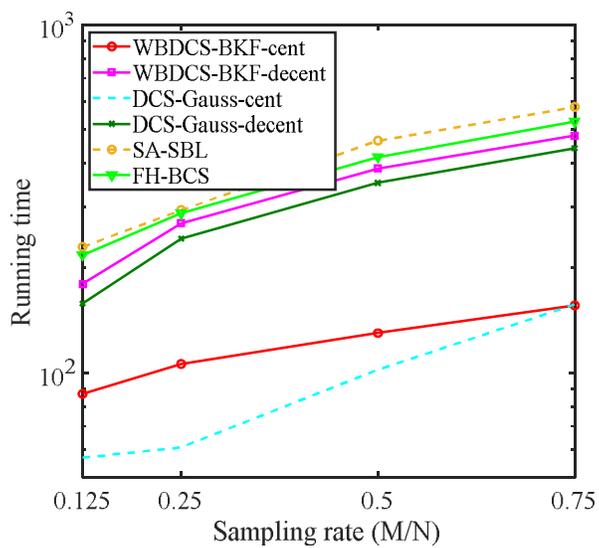

Fig. 12: Running time comparison for the 'Pepper' image shown in fig. 5.

# 7. Conclusion

In this paper, we have proposed a model-based Bayesian DCS algorithm which uses the JSM-1 model and exploits the intra-signal dependencies, as well as the inter-signal dependencies to jointly recover multiple sparse signals. The proposed approach benefits from the intra- and inter-scale dependencies among the wavelet coefficients and utilizes the sparse signal structures and hence, goes beyond the conventional DCS recovery algorithms. Moreover, we have used the Bessel K-form as the prior distribution for the elements of sparse signal. The BKF has a heavier tails than the Gaussian pdf and hence, can model the sparsity of the coefficients better than the Gaussian. The VB inference is used based on the statistical structure of the sparse signals and the BKF prior, and a closed-form solution for model parameters are obtained. Experimental results showed that the proposed model-based DCS algorithm significantly outperforms existing approaches in terms of recovery performance and convergence properties.